\documentclass[conference, 9pt]{IEEEtran}
\IEEEoverridecommandlockouts
\usepackage[utf8]{inputenc}
\usepackage{amsmath,amssymb,amsfonts}
\usepackage{graphicx}
\usepackage{cite}
\usepackage{url}
\usepackage{booktabs}
\usepackage{multirow}
\usepackage{listings} 
\usepackage{algorithm} 
\usepackage{algpseudocode}
\usepackage{hyperref} 
\usepackage{subcaption} 
\usepackage{tikz}

\newcommand{\circleNumber}[1]{%
  \begin{tikzpicture}[baseline=(char.base)]
    \node[shape=circle, fill=black, inner sep=2pt] (char) {\textcolor{white}{#1}};
  \end{tikzpicture}
}

\begin{document}

\title{Early Detection of Hardware Trojans Using Neural Controlled Differential Equations and Analysis of Power Traces}

\author{\IEEEauthorblockN{Hasala Senevirathne}
\IEEEauthorblockA{\textit{Dept. of Computer Eng. \& Computer Sci.} \\
\textit{California State University, Long Beach}\\
Long Beach, CA, United States\\
hasala.senevirathne01@student.csulb.edu}
\and
\IEEEauthorblockN{Rahul Vishwakarma}
\IEEEauthorblockA{\textit{Dept. of Computer Eng. \& Computer Sci.} \\
\textit{California State University, Long Beach}\\
Long Beach, CA, United States \\
rahuldeo.vishwakarma01@student.csulb.edu}
\and
\IEEEauthorblockN{Amin Rezaei}
\IEEEauthorblockA{\textit{Dept. of Computer Eng. \& Computer Sci.} \\
\textit{California State University, Long Beach}\\
Long Beach, CA, United States \\
amin.rezaei@csulb.edu }
}

\maketitle

\begin{abstract}
 Evolving Hardware Trojans pose a serious threat to modern digital systems by evading traditional detection through stealthy, adaptive behavior. Even recent methods that leverage advances in machine learning can only detect them after activation, leaving a critical window for potential security breaches. To address this gap, we propose a novel approach for hardware Trojan detection and prediction using Neural Controlled Differential Equations (NCDEs) and analysis of power traces. Our method leverages an NCDE model trained exclusively on Trojan-free data to learn nominal power behavior, combined with a Linear Discriminant Analysis (LDA) classifier calibrated on labeled data, to distinguish between three scenarios: no Trojan, dormant Trojan, and active Trojan. Our method uses a sliding window to process side-channel measurements, enabling detection of subtle power consumption deviations that indicate Trojan presence, even when dormant. Experimental results demonstrate that the proposed NCDE-based method achieves superior accuracy compared to traditional machine learning approaches, with the additional advantage of handling dormant Trojans above a sensitivity threshold. We validate our approach on standard hardware Trojan benchmarks, showing robust detection and prediction performance. \\

\end{abstract}

\begin{IEEEkeywords}
Hardware Trojan Detection, Neural Controlled Differential Equations, Power Side-Channel Analysis
\end{IEEEkeywords}

\section{Introduction}
Evolving Hardware Trojans (HTs) threaten electronic systems by altering Integrated Circuits (ICs), leading to unauthorized access, data leakage, or system failure~\cite{xue2020ten}. Their stealthy and adaptive nature allows HTs to evade traditional detection methods, making them a key concern in hardware security research. Traditional static detection methods, such as static analysis~\cite{frick2015HT, Cai2023-HTStatic} and logic testing~\cite{mishra2021testing}, struggle with scalability and are impractical for identifying Trojans with rare activation triggers. As ICs grow in size and complexity, exhaustive testing becomes too complex. Dynamic detection methods based on neural networks have been explored~\cite{lao2019htnn}, but models like Recurrent Neural Networks (RNNs)~\cite{li2019htrnn} require significant computational resources and may not capture the irregular, high-dimensional side-channel signals, such as power consumption and electromagnetic emissions, needed to detect Trojans at run-time.

A particularly challenging aspect of HT detection is identifying not only when a Trojan is actively triggered but also when it is present yet dormant in the circuit. This capability is essential for proactive security measures, as it allows for the prediction of malicious hardware before it can cause harm. Traditional detection methods often focus solely on identifying active Trojans, missing the opportunity to detect dormant threats~\cite{karri2010survey, Bhunia2021HTDetection, Hepp2022HTDetection, Pagliarini2022HTDetection, Vishwakarma2023HTDetection, wang2023secure, Fujimoto2023HTDetection, Vishwakarma2024HTDetection, Vishwakarma2025HTDetection, Fujimoto2024HTDetection, Yasaei2025HTDetection, Shiomi2025HTDetection, Maynard2024HDMitigation, Faruque2025DSHTDetection, Hoque2025HTLLMDetection, Yang2025HTDetection, Yu2025HTDetection, Hong2026HTDetection}. 

Neural Controlled Differential Equations (NCDEs) have recently gained attention for their ability to model continuous-time dynamics in irregular time-series data~\cite{kidger2020neural}. NCDEs extend neural networks into continuous time, effectively capturing intricate temporal dependencies and accommodating irregularly sampled observations. This makes them particularly well-suited for processing side-channel signals that are inherently irregular and high-dimensional. We believe that by modeling the temporal evolution of these signals, NCDEs have the potential to identify subtle deviations from normal behavior, indicating the presence of dormant Trojans even when they are not actively triggered.

In this paper, we investigate the potential of using NCDEs combined with a sliding window mechanism to analyze power consumption traces, enabling not only the detection of HTs upon activation but also the prediction of their presence while they remain dormant. Our method leverages an NCDE model trained on Trojan-free data, combined with an LDA classifier calibrated on labeled traces, to distinguish between three scenarios: (1) No Trojan presented, (2) Trojan presented but dormant, and (3) Trojan presented and active. This three-way classification provides a more nuanced view of system security than traditional binary classification approaches. Our main contributions are as follows:

\begin{itemize}
    \item Proposing a novel three-way classification approach for detection and prediction of HTs, leveraging NCDEs to effectively model power side-channel signals;
    \item Exploring a threshold at which the model can reliably catch the presence of HTs, even while they remain dormant;    
    \item Implementing an efficient sliding window approach for processing power trace data, enabling systematic Trojan detection and state classification.
\end{itemize}

The rest of the paper is organized as follows: Section~\ref{sec:related} reviews related works on HT detection and time series. Section~\ref{sec:background} covers the preliminaries on HTs, side-channel analysis, and NCDEs. Section~\ref{sec:proposed} outlines our methodology for early Trojan detection using NCDEs. Section~\ref{sec:results} presents the experimental setup, results, and analysis. Finally, Section~\ref{sec:conclusion} concludes the paper and discusses future works.

\section{Related Works}\label{sec:related}
\begin{figure*}[!t]
    \centering
    \includegraphics[width=\textwidth]{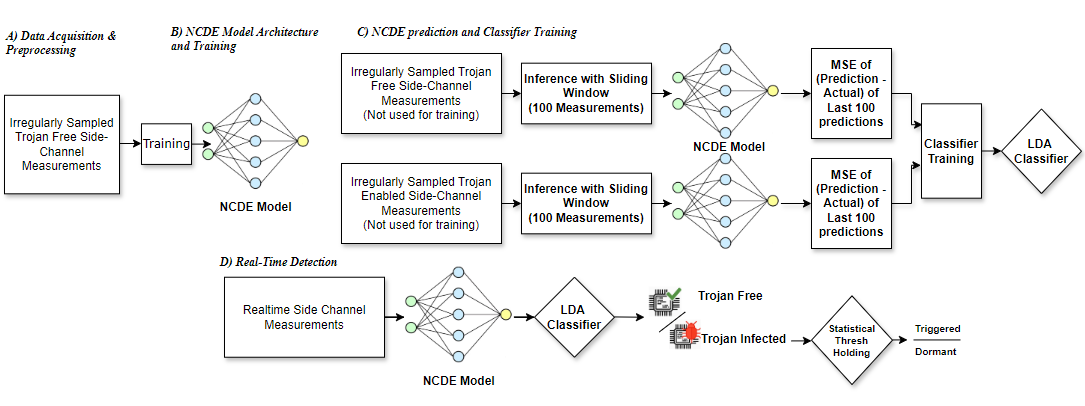}
    \caption{\textbf{HOODOO}: NCDE-based Hardware Trojan Detection and Prediction Framework}
    \label{fig:HOODOO}
\end{figure*}

\subsection{Hardware Trojan Detection}
HT detection approaches are broadly categorized into pre-silicon~\cite{ma2024presilicon} and post-silicon~\cite{puschner2023redteam} methods. Pre-silicon techniques focus on design-time verification, while post-silicon methods involve physical inspection and side-channel analysis~\cite{mosavirik2023silicon}. A comprehensive survey~\cite{tehranipoor2010survey} underscores the significant challenges in detecting dormant hardware Trojans, a key focus of this work.

Side-channel analysis has proven effective. Path delay fingerprinting~\cite{jin2008hardware} identifies timing anomalies but requires extensive characterization. Power consumption analysis~\cite{agrawal2007trojan} compares measurements with golden models but struggles with process variations. ML approaches include Multi-Layer Neural Networks (MLNNs)~\cite{hasegawa2017hardware} and Long Short-Term Memories (LSTMs)~\cite{nasr2024siamese} for analyzing power traces, though they require extensive training data and may not capture continuous-time dynamics. A brain-inspired model known as Hierarchical Temporal Memory (HTM) is proposed for HT detection and is designed to be resilient to natural variations in side-channel measurements \cite{Faezi2021HTDetection}. Recently, the application of Large Language Models (LLMs) to hardware Trojan (HT) detection has also been explored, both with and without the inclusion of contextual supplementary information~\cite{Tehranipour2024HTDetection}. However, these efforts are still in the early stages, and further research is needed to evaluate the effectiveness of LLMs in detecting zero-day HTs.

\subsection{Time Series Analysis}
Neural Ordinary Differential Equations (Neural ODEs)~\cite{chen2018neural} introduced continuous-time generalization of residual networks, representing hidden state dynamics as ODEs for flexible continuous-time modeling. Building on Neural ODEs, NCDEs~\cite{kidger2020neural} were proposed for irregularly sampled time series, parameterizing the ODE vector field using neural networks with control paths from observed data. The NCDE framework has been extended for noisy/partially observed series with improved gradient computation and stability~\cite{morrill2021neural}, crucial for noisy side-channel analysis. GRU-ODE-Bayes~\cite{debrouwer2019gru} combines gated recurrent units with neural ODEs for variable sampling rates in monitoring scenarios. While these approaches show promise across domains, their hardware security and HT detection applications remain unexplored. Our work bridges this gap by adapting NCDEs to side-channel-based HT detection problem.

Our work leverages the continuous-time modeling capabilities of NCDEs within a sliding window framework, enabling high detection (when active) and prediction (when dormant) accuracy of HTs, without requiring hardware modifications.

\section{Preliminaries}\label{sec:background}

\subsection{Hardware Trojans}
HTs are malicious IC modifications that modify functionality, leak information, or cause system failure~\cite{xiao2016hardware}. They consist of a trigger and a payload, with the trigger activating the Trojan under specific hard-to-detect conditions and the payload executing the malicious function. 
Our approach identifies three scenarios: (1)~\textbf{No Trojan}: the circuit is clean; (2)~\textbf{Dormant Trojan}: a Trojan is physically present but its trigger condition has not been met, its presence may still create a measurable, albeit subtle, side-channel footprint; and (3)~\textbf{Active Trojan}: the Trojan is triggered and executes its payload, producing a pronounced deviation in power consumption.

\subsection{Side-Channel Analysis}
Side-channel analysis leverages physical characteristics like power consumption, timing, or electromagnetic emissions to infer circuit operations~\cite{kocher1999differential}. Power analysis is widely used for HT detection, measuring IC power consumption and comparing it with expected profiles~\cite{tehranipoor2010survey}. Deviations may indicate Trojan presence. Its advantages include non-invasiveness, high sensitivity to behavioral changes, and broad applicability across circuit types.

However, process variations can mask Trojan-induced changes, environmental noise reduces detection sensitivity, and modern ICs produce intricate signatures where irregular sampling may miss critical events. NCDEs address these challenges by effectively modeling continuous-time power signal dynamics and handling irregularly sampled or noisy data.

\subsection{Neural Controlled Differential Equations}
Unlike traditional RNNs that operate in discrete time steps, NCDEs model the evolution of hidden states as a continuous-time process controlled by the input data. Mathematically, an NCDE is defined as:
\begin{equation}
h(t) = h(t_0) + \int_{t_0}^{t} f_\theta(h(s)) \, dX(s),
\label{eq:ncde}
\end{equation}
where $h(t)$ is the hidden state at time $t$, $f_\theta$ is a neural network with parameters $\theta$ governing the hidden state dynamics, $X(t)$ is a continuous interpolation of the input data, and $dX(s)$ represents the differential of the control path. In practice, input data points are interpolated to create $X(t)$ using cubic splines or Hermite cubic interpolation~\cite{morrill2021online}, and the equation is solved numerically using methods like Runge-Kutta. NCDEs naturally handle irregular sampling, capture intricate temporal dynamics through continuous-time modeling, and offer architectural flexibility---properties well-suited for HT detection using power measurements.

\section{Trojan Detection and Prediction}\label{sec:proposed} 
 
In this section, we present \textbf{HOODOO}, a \textbf{H}ardware Tr\textbf{O}jan detecti\textbf{O}n and pre\textbf{D}icti\textbf{O}n framew\textbf{O}rk using NCDEs shown in Fig. \ref{fig:HOODOO}. The central idea is to train an NCDE model on power trace data obtained from Trojan-free hardware to learn its nominal behavior. We then use this model to identify anomalies in new measurements that may suggest the presence of a Trojan, distinguishing between dormant and active states. 
\textbf{HOODOO} consists of the following components: \textbf{\circleNumber{A}Data Acquisition and Preprocessing}, which involves collecting power consumption traces from the device under test, followed by normalization and sliding window segmentation of these traces; \textbf{\circleNumber{B}NCDE Model Architecture and Training}, a time-series model trained and optimized on Trojan-free data; \textbf{\circleNumber{C}NCDE Prediction and Classifier Training}, which employs the trained NCDE model and labeled power trace data to train the classifier.
\textbf{\circleNumber{D}Detection and Classification}, which categorizes the device as non-Trojan, dormant Trojan, or active Trojan states, by feeding power trace data to the NCDE model and then to the classifier.

\subsection{Data Acquisition and Preprocessing}
We employ a publicly accessible online dataset~\cite{dataset_rozhin1} with high-resolution power consumption traces from hardware platforms in different operational states. To process power trace data, we use a sliding window technique with a buffer of $W$ samples. The window advances by removing old and adding new measurements. For each window, we apply three pre-processing steps:
\begin{enumerate}
\item \textbf{Normalization}: Subtract mean and divide by standard deviation from training data: $x_{\mathrm{normalized}} = \frac{x - \mu_{\mathrm{train}}}{\sigma_{\mathrm{train}}}$, where $\mu_{\mathrm{train}}$ and $\sigma_{\mathrm{train}}$ are training dataset statistics. This ensures zero mean and unit variance for stable training.
\item \textbf{Time Step Assignment}: Assign equidistant time steps in $[0, 1]$: $t_i = \frac{i}{W-1}$ for $i = 0, 1, \ldots, W-1$.
\item \textbf{Interpolation Coefficient Computation}: Compute cubic spline or Hermite cubic interpolation coefficients to represent the data as a continuous path.
\end{enumerate}
The normalized window and interpolation coefficients are fed to the NCDE model for processing.

\subsection{NCDE Model Architecture and Training}
Our NCDE model architecture consists of three main components:
\begin{enumerate}
\item \textbf{Initial Mapping}: A linear layer that maps the initial data point in the window to the initial hidden state: $h(t_0) = \mathrm{Linear}(X(t_0))$.
\item \textbf{CDE Function}: A neural network that defines the dynamics of the hidden state: $f_\theta(h(t)) = \mathrm{NeuralNetwork}(h(t))$.
\item \textbf{Readout Layer}: A linear layer that maps the final hidden state to the predicted output: $\hat{y} = \mathrm{Readout}(h(t))$.
\end{enumerate}

\begin{algorithm} [!t]
\caption{NCDE Function}
\label{Alg:CDE-Func}
\begin{algorithmic}[1]
\Function{CDEFunc}{$h$, $input\_dim$, $hidden\_dim$}
    \State $batch\_size \gets$ size of first dimension of $h$
    \State $x \gets \mathbf{ones}(batch\_size, input\_dim)$
    \State $xz \gets \text{concatenate}(x, h)$
    \State $hidden \gets \text{Linear}(xz, hidden\_dim \times 2)$
    \State $hidden \gets \text{GELU}(hidden)$
    \State $z \gets \text{Linear}(hidden, input\_dim \times hidden\_dim)$
    \State \Return reshape($z$, [$batch\_size$, $hidden\_dim$, $input\_dim$])
\EndFunction
\end{algorithmic}
\end{algorithm}

Algorithm~\ref{Alg:CDE-Func} parametrizes the CDE vector field: it concatenates a bias vector of ones with the hidden state $h$, processes this through a two-layer MLP with GELU activation, and reshapes the output for CDE integration.

\begin{algorithm} [!t]
\caption{NCDE Model Forward Pass}
\label{Alg:NCDE-Forw}
\begin{algorithmic}[1]
\Function{NCDEForward}{$coeffs$, $input\_dim$, $hidden\_dim$, $output\_dim$}
    \State \# Create continuous path
    \State $X \gets \text{CubicSpline}(coeffs)$
    \State \# Initial point
    \State $X_0 \gets X.\text{evaluate}(X.interval[0])$
    \State \# Initial hidden state
    \State $z_0 \gets \text{Linear}(X_0, hidden\_dim)$
    \State \# Solve differential equation with numerical method
    \State $h_T \gets \text{CDEint}(X, z_0, \text{CDEFunc}, X.grid\_points,$
    \State $\quad method=\text{`rk4'})$
    \State \# Apply readout
    \State $output \gets \text{Linear}(h_T[:, -1, :], output\_dim)$
    \State \Return $output$
\EndFunction
\end{algorithmic}
\end{algorithm}
 
Algorithm~\ref{Alg:NCDE-Forw} implements the forward pass: it creates a continuous path via cubic spline interpolation, initializes the hidden state through a linear mapping, solves the CDE using 4th-order Runge-Kutta, and extracts the final hidden state for prediction through a readout layer.

\begin{algorithm} [!t]
\caption{NCDE-based Trojan Detection}
\label{Alg:HT-Detection}
\begin{algorithmic}[1]
\Function{DetectTrojan}{$model$, $power\_trace$, $W$, $b_{LDA}$, $T_{triggered}$}
    \State $errors \gets []$
    \For{$i \gets 0$ to $\text{len}(power\_trace) - W - 1$}
        \State $window \gets power\_trace[i:i+W]$
        \State $norm\_window \gets \text{Normalize}(window)$
        \State $coeffs \gets \text{ComputeCoefficients}(norm\_window)$
        \State \# NCDE predicts the next sample
        \State $\hat{y} \gets model(coeffs)$
        \State \# Actual next sample
        \State $y \gets power\_trace[i + W]$
        \State \# Squared prediction error
        \State $errors.\text{append}((\hat{y} - y)^2)$
    \EndFor
    \State $MSE \gets \text{mean}(errors)$
    \If{$MSE \leq b_{LDA}$}
        \State \Return ``No Trojan Presented''
    \ElsIf{$MSE \leq T_{triggered}$}
        \State \Return ``Trojan Presented but Dormant''
    \Else
        \State \Return ``Trojan Presented and Active''
    \EndIf
\EndFunction
\end{algorithmic}
\end{algorithm}

Algorithm~\ref{Alg:HT-Detection} implements the three-state classification: for each sliding window of $W$ samples, it predicts the next power sample via the NCDE model, computes the squared prediction error, and classifies the aggregate MSE using the LDA-derived threshold $b_{LDA}$ and the active Trojan threshold $T_{triggered}$.

We train our NCDE model exclusively on Trojan-free power traces to predict the next power sample given a window of $W$ preceding samples. By learning normal circuit behavior, the model produces higher prediction errors on Trojan-infected circuits, which the LDA classifier leverages for detection. Training optimizes the NCDE parameters to minimize MSE between predicted and actual next power values:
\begin{equation}
\mathcal{L}(\theta) = \frac{1}{N} \sum_{i=0}^{N-1} \bigl(y_i - \hat{y}_i\bigr)^2,
\end{equation}
where $y_i$ is the actual next power value and $\hat{y}_i$ is the predicted value.

We also employ several optimization techniques to improve training efficiency: 
\begin{enumerate}
\item \textbf{AdamW optimizer} with weight decay for regularization.
\item \textbf{OneCycleLR scheduler} for dynamic learning rate adjustment.
\item \textbf{Mixed-precision training} for faster computation on compatible GPUs.
\item \textbf{Gradient scaling} to prevent underflow in mixed-precision training.
\end{enumerate}

\subsection{NCDE Prediction and Classifier Training}
During inference, for each window of $W$ samples starting at index $i$, the NCDE model predicts the next value $\hat{y}_{i+W}$, and the squared prediction error is computed:
\begin{equation}
e_i = (\hat{y}_{i+W} - y_{i+W})^2.
\label{eq:pred_error}
\end{equation}
These errors are aggregated over $N$ windows to obtain the MSE:
\begin{equation}
\mathrm{MSE} = \frac{1}{N} \sum_{i=0}^{N-1} e_i.
\label{eq:mse_calc}
\end{equation}
A higher MSE indicates greater deviation from nominal behavior. We employ Linear Discriminant Analysis (LDA) for classification, where the feature is the MSE value and the classes are Trojan-free ($C_0$) and Trojan-infected ($C_1$). While the NCDE model trains on Trojan-free data only, the LDA classifier requires a small labeled calibration set from both classes. LDA maximizes the between-class to within-class variance ratio, determining an optimal boundary $b_{LDA}$. Classification of a new trace proceeds as:
\begin{equation}
\text{Classification} =
\begin{cases}
\text{Trojan-free} & \text{if } \mathrm{MSE}_{new} \leq b_{LDA} \\
\text{Trojan Dormant} & \text{if } b_{LDA} < \mathrm{MSE}_{new} \leq T_{triggered} \\
\text{Trojan Active} & \text{if } \mathrm{MSE}_{new} > T_{triggered}
\end{cases}
\label{eq:lda_classification}
\end{equation}
where $b_{LDA}$ is the optimal threshold from LDA. The second threshold $T_{triggered}$ distinguishes dormant from active states; however, its calibration requires labeled traces with known trigger timestamps, which are unavailable in the current benchmark. As this work focuses on dormant Trojan detection, the experimental evaluation centers on $b_{LDA}$, while the dormant-versus-active distinction is validated through noise injection at varying intensity levels.

\section{Experimental Results}\label{sec:results}

\subsection{Experimental Setup}
We used the publicly available hardware Trojan power side-channel dataset~\cite{dataset_rozhin1}, which contains power traces measured from physical hardware using a Sakura-G FPGA board and a Tektronix TDS2022C oscilloscope. The dataset includes TrustHub benchmarks with two base circuits: an AES-128 cryptography core (9707 LUTs) and an RS232 UART serial communication circuit, infected with six Trojan variants: AES-T500, AES-T600, AES-T700, AES-T800, AES-T1600, and RS232-T100. These Trojans range from 0.29\% to 4.46\% of the base circuit area and implement diverse payloads including denial of service, secret key leakage through leakage current and covert channels, and RF transmission. Power traces were collected under two conditions per benchmark: HT inactive (dormant) and HT activated (triggered), along with Trojan-free baseline traces. Each category contains 10,000 traces of 2,500 samples each. Process variation was addressed in the original dataset by collecting traces from two separate Sakura-G boards.

Power traces were segmented into sliding windows of $W=50$ samples with a stride of 1 sample. Trojan-free data was split 80/20 for training/validation using fixed-seed random splitting. The LDA classifier was calibrated using MSE values derived from both Trojan-free and Trojan-infected traces. The NCDE model was implemented in PyTorch with configurations in Tables~\ref{tab:ncde_model_config} and~\ref{tab:training_config}. The experiments ran on a 12-core Intel Xeon Silver 4310 processor with 16GB of DDR4 RAM and an NVIDIA 64GB A16 GPU.

\begin{table}[!t]
\centering
\caption{NCDE Model Configuration}
\label{tab:ncde_model_config}
\begin{tabular}{p{3.2cm}|p{4.8cm}}
\hline
\textbf{Parameter} & \textbf{Value} \\
\hline
\hline
Input channels & 2 (time, power) \\
Hidden channels & 64 \\
Output channels & 1 (predicted power) \\
CDE function & 2-layer MLP, GELU, dropout 0.1 \\
Numerical solver & RK4, step size 0.2 \\
Batch size & 256 (GPU), 64 (CPU) \\
Epochs & 30 \\
\hline
\end{tabular}
\end{table}

\begin{table}[!t]
\centering
\caption{Training Configuration}
\label{tab:training_config}
\begin{tabular}{p{3.2cm}|p{4.8cm}}
\hline
\textbf{Parameter} & \textbf{Value} \\
\hline
\hline
Optimizer & AdamW, LR $1\times10^{-3}$, WD $1\times10^{-4}$ \\
LR scheduler & OneCycleLR, max LR $1\times10^{-3}$ \\
Loss function & Mean Squared Error (MSE) \\
Mixed precision & Enabled (if GPU supports) \\
\hline
\end{tabular}
\end{table}

\subsection{Evaluation Methodology}
We evaluated our approach using the following methodology:
\begin{enumerate}
\item \textbf{Train-Test Split}: The NCDE model was trained exclusively on Trojan-free power traces. Testing was performed on separate sets for each category (no Trojan, dormant, active).
\item \textbf{Error Metrics}: For each sliding window of $W=50$ samples, the NCDE model predicts the next power sample, and the squared prediction error is computed. The MSE is aggregated across all windows in a trace.
\item \textbf{Threshold Determination}: An LDA classifier was trained on MSE values from both Trojan-free and Trojan-infected traces to determine the optimal decision boundary $b_{LDA}$ for separating clean and Trojan-infected states. The dormant-versus-active distinction is evaluated through noise injection at varying intensity levels (1--5\% of peak amplitude), as the benchmark dataset does not provide explicit trigger timestamps for calibrating $T_{triggered}$.
\item \textbf{Classification}: Each trace is classified based on its MSE:
$\mathrm{MSE} \leq b_{LDA} \Rightarrow$ No Trojan;
$\mathrm{MSE} > b_{LDA} \Rightarrow$ Trojan Detected (dormant or active, distinguished by deviation magnitude).
\end{enumerate}

\begin{figure}[!t]
    \centering

    \begin{subfigure}[b]{0.95\columnwidth}
        \centering
        \includegraphics[width=\linewidth]{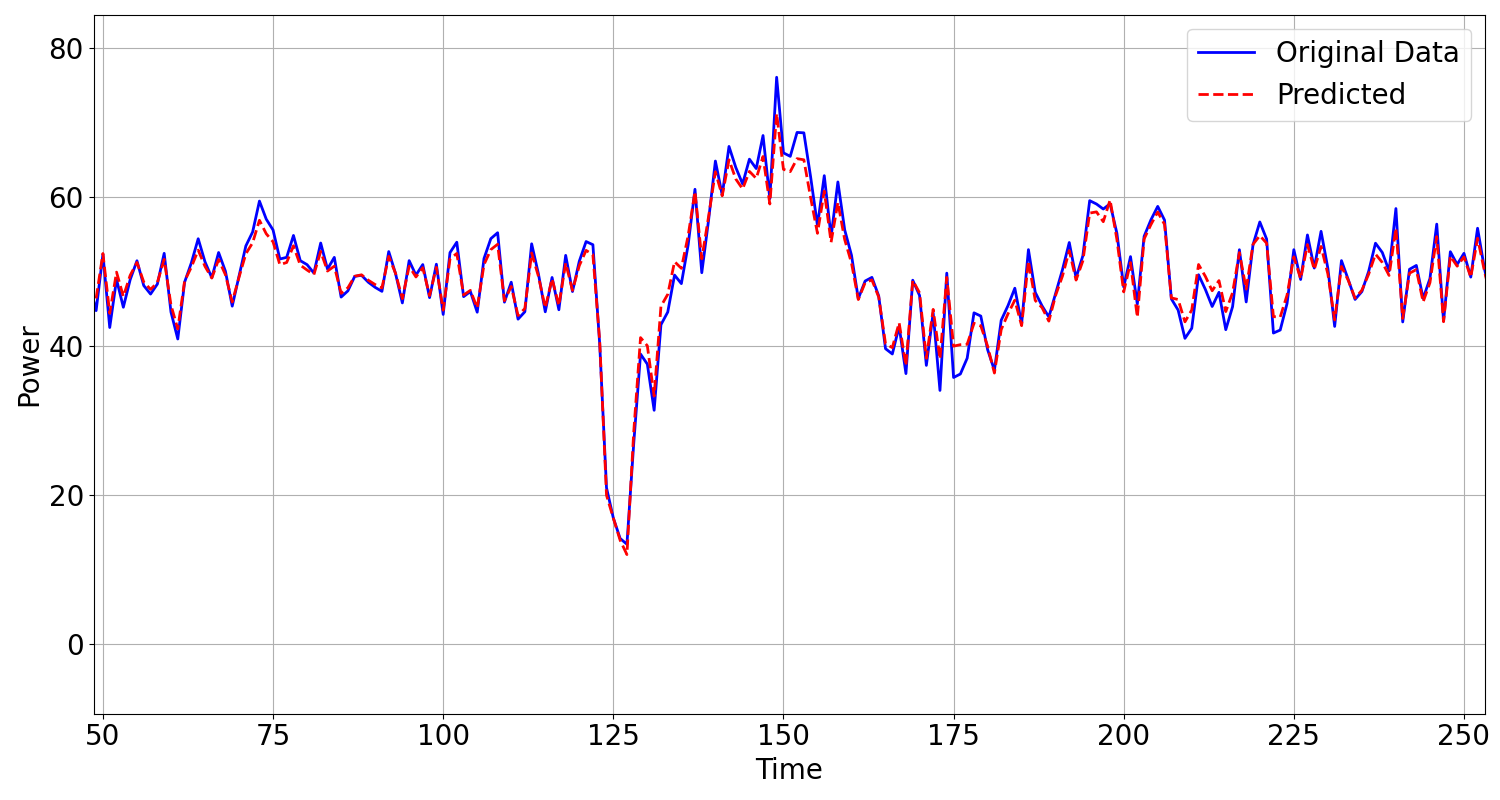}
        \caption{Trojan Disabled}
        \label{fig:trojan_disabled}
    \end{subfigure}
    \hfill
    \begin{subfigure}[b]{0.95\columnwidth}
        \centering
        \includegraphics[width=\linewidth]{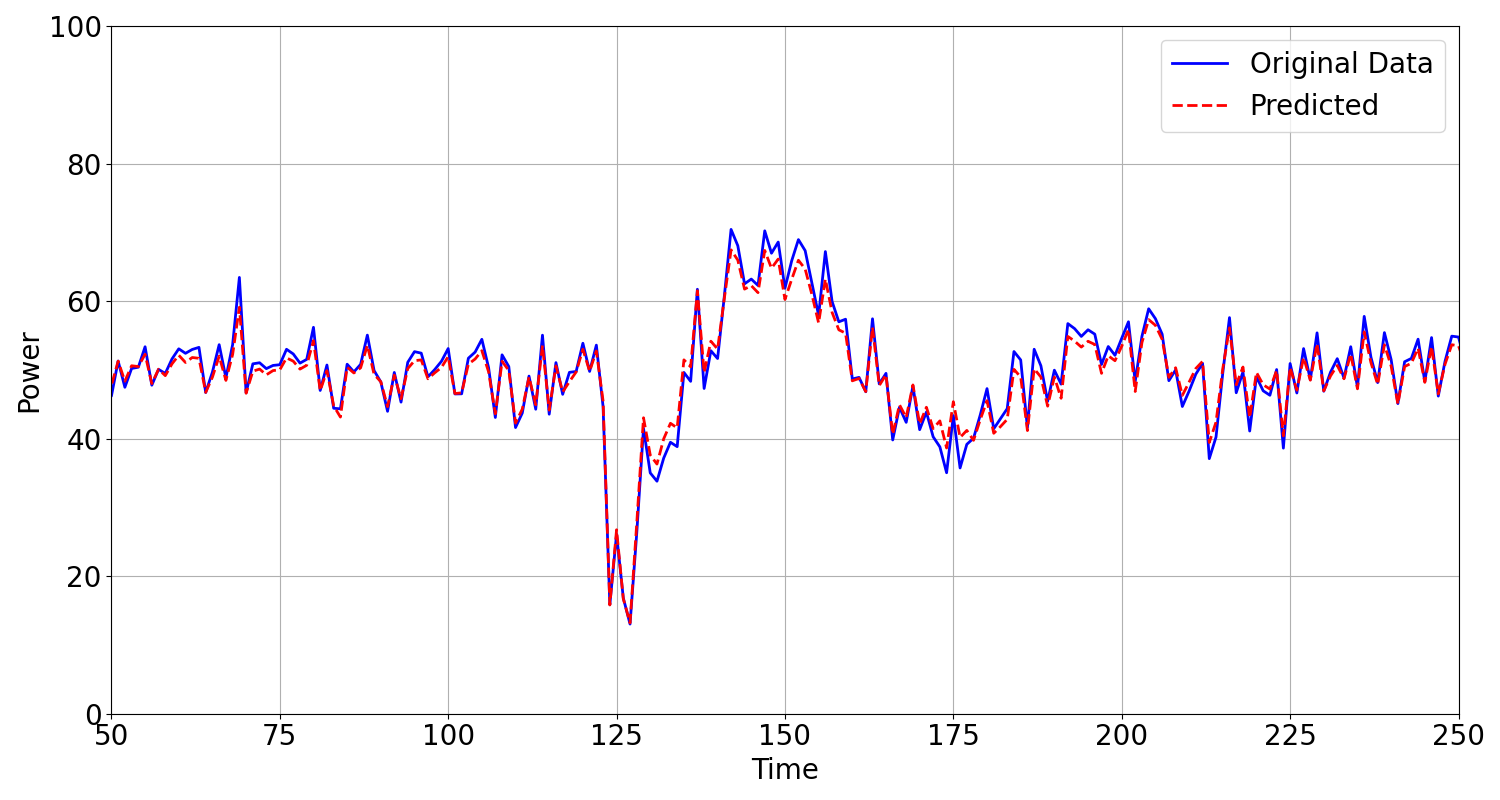}
        \caption{Trojan Dormant}
        \label{fig:trojan_dormant}
    \end{subfigure}
    \hfill
    \begin{subfigure}[b]{0.95\columnwidth}
        \centering
        \includegraphics[width=\linewidth]{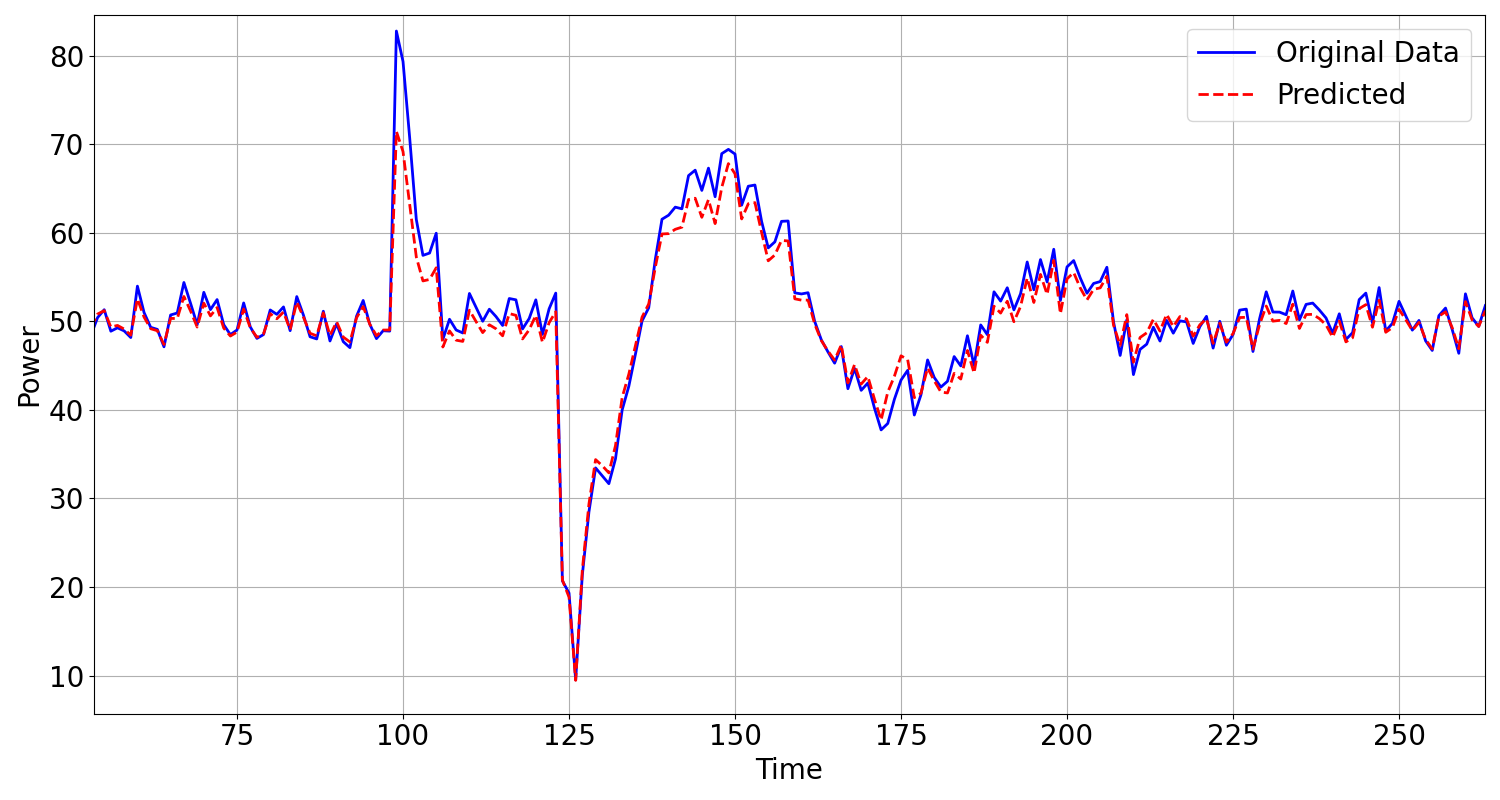}
        \caption{Trojan Triggered}
        \label{fig:trojan_triggered}
    \end{subfigure}

    \caption{NCDE Predictions under Different HT conditions}
    \label{fig:comparison}
\end{figure}

\subsection{NCDE Model Performance}
Fig.~\ref{fig:trojan_disabled} validates the NCDE model's baseline performance in predicting nominal circuit behavior in the absence of Trojans. Fig.~\ref{fig:trojan_dormant} quantifies the prediction residuals observed when dormant Trojans introduce subtle perturbations in power consumption patterns. Fig.~\ref{fig:trojan_triggered} exhibits the model's response to active Trojan payloads, contrasting predicted power traces with actual measurements to highlight the detectable deviations. As shown in the figures, the NCDE model demonstrates strong performance in predicting continuous values.

\subsection{HT Detection Performance}
\begin{figure}[!t]
     \centering
        \includegraphics[width=\linewidth]{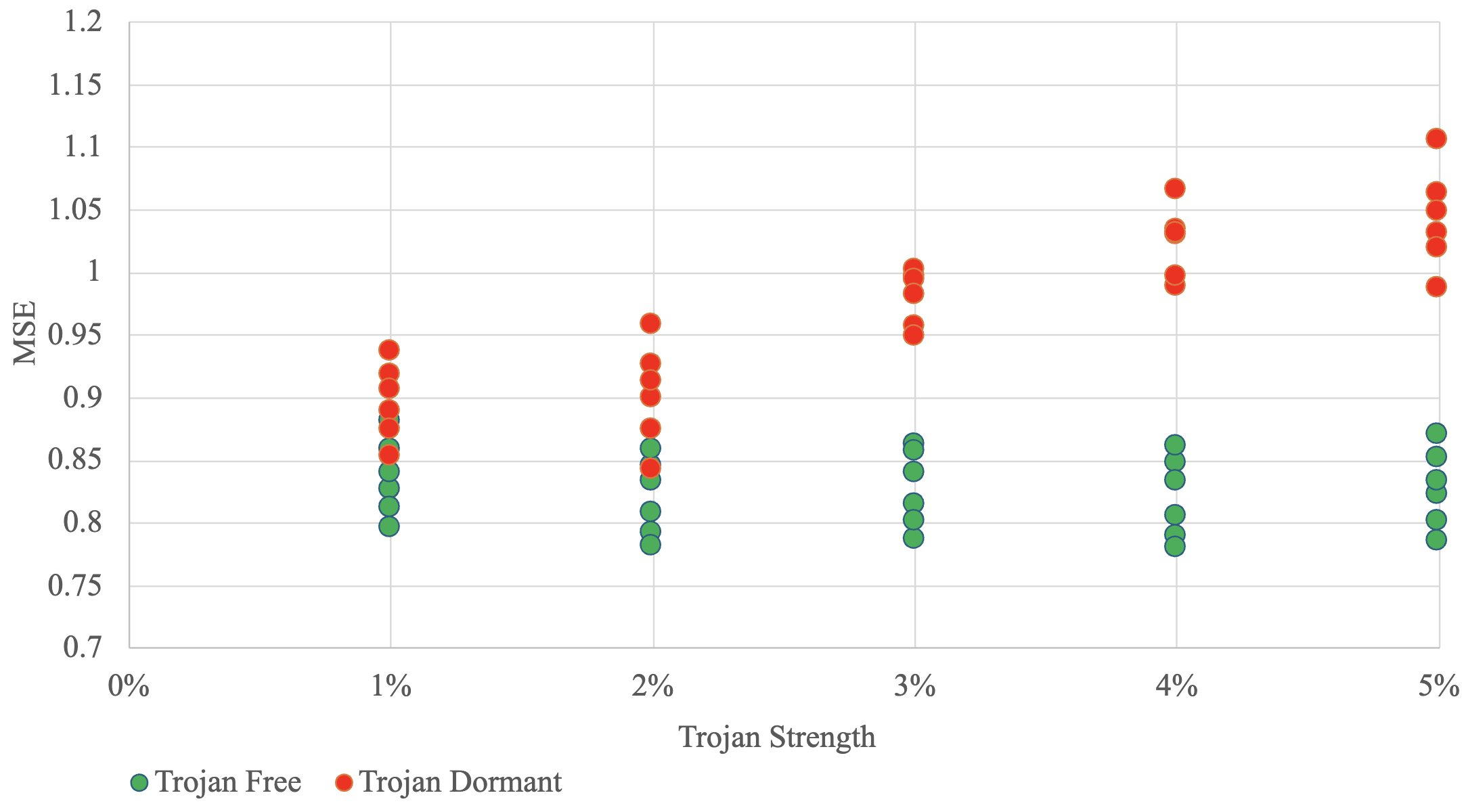}
        \caption{Detection Sensitivity Thresholds of HTs}
        \label{fig:trojanstrength}
\end{figure}

To evaluate detection sensitivity, we injected normally distributed noise at varying levels (1--5\% of peak power trace amplitude) into Trojan-free traces, simulating the subtle power deviations that stealthy hardware Trojans introduce to evade detection~\cite{Omidi2024EvasiveHardwareTrojan}. This provides a controlled evaluation scenario for assessing the minimum perturbation level at which the model can reliably distinguish Trojan-affected behavior from nominal operation. As shown in Fig.~\ref{fig:trojanstrength}, our NCDE-based method successfully detected Trojans at noise levels $\geq$3\% of peak power value. Below this 3\% threshold, the model's ability to distinguish dormant Trojans from normal behavior degraded, indicating a critical detection boundary where Trojans inducing variations $<$3\% can evade detection, highlighting the need for enhanced sensitivity.

\subsection{Comparison with State-of-the-Art Methods}
Table~\ref{tab:related} compares our approach with recent ML-based HT detection methods~\cite{hasegawa2017hardware, nasr2024siamese, Faezi2021HTDetection, Tehranipour2024HTDetection}. The compared methods use different input modalities: MLNN~\cite{hasegawa2017hardware} operates on gate-level netlists, LSTM~\cite{nasr2024siamese} and HTM~\cite{Faezi2021HTDetection} use power traces, and the LLM-based method~\cite{Tehranipour2024HTDetection} operates at the RTL/netlist level. Our method achieves competitive or superior active Trojan detection accuracy and is the only approach offering three-state classification with dormant detection capability.

\subsection{Limitations and Challenges}
Our detection assumes dormant Trojans produce a measurable side-channel footprint; Trojans with power variations below 3\% of peak amplitude may evade detection. Also, the sensitivity analysis uses Gaussian noise as a proxy; real dormant Trojans may exhibit structured, non-Gaussian signatures. Finally, MSE is the sole anomaly feature; richer residual features from NCDE latent states could improve sensitivity.

\begin{table}[!t]
\centering
\caption{Comparison of HOODOO with Existing Methods}
\label{tab:related}
\setlength{\tabcolsep}{2pt}
\begin{tabular}{p{2cm}ccc}
\toprule
\textbf{Method} & \textbf{3-State Class?} & \textbf{Dormant Acc. (\%)} & \textbf{Active Acc. (\%)} \\
\midrule
\midrule
MLNN~\cite{hasegawa2017hardware} & No & N/A & 85.0 \\
\midrule
LSTM~\cite{nasr2024siamese} & No & N/A & 86.8 \\
\midrule
HTM~\cite{Faezi2021HTDetection} & No & N/A & 92.2 \\
\midrule
\multirow{2}{*}{LLM (GPT-4) \cite{Tehranipour2024HTDetection}} & \multirow{2}{*}{No} & \multirow{2}{*}{N/A} & Context-Free: 81.0 \\
& & & Contextual: 91.7 \\
\midrule
\multirow{5}{*}{NCDE (Ours)}  & \multirow{5}{*}{Yes} & 1\% Thr: 55.7 & 1\% Thr: 92.4\\
 &  & 2\% Thr: 62.5 & 2\% Thr: 94.6 \\
 &  & 3\% Thr: 80.2 & 3\% Thr: 95.4 \\
 &  & 4\% Thr: 88.3 & 4\% Thr: 99.3 \\
 &  & 5\% Thr: 92.2 & 5\% Thr: 100.0 \\
\bottomrule
\end{tabular}
\end{table}

\section{Conclusion}\label{sec:conclusion}
In this paper, we presented a novel approach for HT detection and prediction, employing NCDEs. The proposed approach enables effective three-state classification, distinguishing between no Trojan, dormant Trojan, and active Trojan conditions to support proactive security measures. Experimental validation demonstrates superior performance relative to conventional machine learning techniques. The core framework, coupling NCDEs with temporal analysis, shows significant potential for broader applications within hardware security and reliability, including but not limited to, side-channel attack mitigation, device aging monitoring, and fault detection paradigms. 

Future research directions include integrating multi-modal side-channel data to improve the sensitivity of dormant Trojan detection, exploring richer anomaly features from NCDE latent states beyond MSE, evaluating generalization across diverse chip designs and process variation conditions, and applying transfer learning techniques to reduce training data requirements for new hardware designs.

\section*{Acknowledgment}
This material is based upon work supported by the National Science Foundation under Award No. 2245247.

\bibliographystyle{IEEEtran}

\footnotesize
\begin{thebibliography}{99}
\footnotesize

\bibitem{xue2020ten}
 M. Xue, C. Gu, W. Liu, S. Yu, and M. O'Neill, ``Ten years of hardware Trojans: A survey from the attacker's perspective,'' In \emph{IET Computers \& Digital Techniques}, vol.~14, pp. 231--246, 2020.

 \bibitem{frick2015HT}
 J. Francq and F. Frick, ``Introduction to hardware Trojan detection methods,'' In \emph{Design, Automation \& Test in Europe Conference \& Exhibition (DATE)}, pp. 770--775, 2015.

\bibitem{Cai2023-HTStatic}
H. Wang, Q. Zhou, and Y. Cai, ``Static probability analysis guided RTL hardware Trojan test generation,'' In \emph{Asia and South Pacific Design Automation Conference}, pp. 510--515, 2023.

 \bibitem{mishra2021testing}
 Z. Pan and P. Mishra, ``Automated test generation for hardware Trojan detection using reinforcement learning,'' In \emph{Asia and South Pacific Design Automation Conference (ASP-DAC)}, pp. 408--413, 2021. 

\bibitem{lao2019htnn}
J. Clements and Y. Lao, ``Hardware Trojan design on neural networks,'' In \emph{IEEE International Symposium on Circuits and Systems (ISCAS)}, pp. 1-5, 2019.

\bibitem{li2019htrnn}
R. Lu, H. Shen, Y. Su, H. Li, and X. Li, ``GramsDet: Hardware Trojan detection based on recurrent neural network,'' In \emph{IEEE Asian Test Symposium (ATS)}, pp. 111--115, 2019.

\bibitem{karri2010survey}
R. Karri, J. Rajendran, K. Rosenfeld, and M.~Tehranipoor, ``Trustworthy hardware: Identifying and classifying hardware Trojans,'' In \emph{IEEE Computer}, vol.~43, no.~10, pp.~39--46, 2010.

\bibitem{Bhunia2021HTDetection}
S. Yang, P. Chakraborty, and S. Bhunia, ``Side-channel analysis for hardware Trojan detection using machine learning,'' in \emph{IEEE International Test Conference India (ITC India)}, pp. 1--6, 2021.

\bibitem{Hepp2022HTDetection} 
A. Hepp, J. Baehr, and G. Sigl, ``Golden model-free hardware Trojan detection by classification of netlist module graphs,'' In \emph{Design, Automation \& Test in Europe Conference \& Exhibition (DATE)}, pp. 1317--1322, 2022.

\bibitem{Pagliarini2022HTDetection} 
T. Perez and S. Pagliarini, ``A side-channel hardware Trojan in 65nm CMOS with 2$\mu$W  precision and multi-bit leakage capability,'' In \emph{Asia and South Pacific Design Automation Conference (ASP-DAC)}, pp. 9--10, 2022. 

\bibitem{Vishwakarma2023HTDetection}
R. Vishwakarma and A. Rezaei, ``Risk-aware and explainable framework for ensuring guaranteed coverage in evolving hardware Trojan detection,'' In \emph{IEEE/ACM International Conference on Computer Aided Design (ICCAD)}, pp. 1--9, 2023.

\bibitem{wang2023secure}
Z.~Wang, Y.~Xue, and H.~Wang, ``Secure run-time hardware Trojan detection using lightweight analytical models,'' In \emph{IEEE International Symposium on Hardware Oriented Security and Trust (HOST)}, pp.~1--9, 2023.

\bibitem{Fujimoto2023HTDetection}
S. Kaji, D. Fujimoto, and Y. Hayashi, ``Simulation-based approach to generating golden data for PCB-level hardware Trojan detection using capacitive sensor,'' In \emph{IEEE Physical Assurance and Inspection of Electronics (PAINE)}, pp. 1--7, 2023.

\bibitem{Vishwakarma2024HTDetection}
R. Vishwakarma and A. Rezaei, ``Uncertainty-aware hardware Trojan detection using multimodal deep learning,'' In \emph{Design, Automation \& Test in Europe Conference \& Exhibition (DATE)}, pp. 1--6, 2024.

\bibitem{Vishwakarma2025HTDetection}
R. Vishwakarma and A. Rezaei, ``Uncertainty-aware unimodal and multimodal learning for evolving hardware Trojan detection,'' In \emph{Journal of Hardware and Systems Security}, Vol. 9, pp. 1--23, 2025.

\bibitem{Fujimoto2024HTDetection}
K. Abe, D. Fujimoto and Y. Hayashi, ``Fundamental study on detecting hardware Trojans in printed circuit boards using ring oscillators,'' In \emph{International Workshop on the Electromagnetic Compatibility of Integrated Circuits (EMC Compo)}, pp. 1--4, 2024.

\bibitem{Yasaei2025HTDetection}
R. Yasaei, L. Chen, S. -Y. Yu, and M. Abdullah Al Faruque, ``Hardware Trojan detection using graph neural networks,'' In \emph{IEEE Transactions on Computer-Aided Design of Integrated Circuits and Systems}, vol. 44, no. 1, pp. 25--38, 2025.

\bibitem{Shiomi2025HTDetection}
T. Ishikawa, K. Yokooji, Y. Midoh, N. Miura, M. Shintani, and J. Shiomi, ``Hardware Trojan detection by fine-grained power domain partitioning,'' In \emph{Asia and South Pacific Design Automation Conference (ASP-DAC)}, pp. 1257--1263, 2025. 

\bibitem{Maynard2024HDMitigation}
J. Maynard and A. Rezaei, ``Reconfigurable run-time hardware Trojan mitigation for logic-locked circuits,'' In \emph{IEEE 17th Dallas Circuits and Systems Conference (DCAS)}, pp. 1--6, 2024.

\bibitem{Faruque2025DSHTDetection}
L. Chen, Y. Gamal, Y. Li, S. -Y. Yu, I. Alouani, and M. A. A. Faruque, ``DART: Distribution-aware hardware Trojan detection,'' In \emph{IEEE Transactions on Information Forensics and Security}, vol. 20, pp. 9600--9609, 2025.

\bibitem{Hoque2025HTLLMDetection}
R. Kumar Kundu, K. Khalil, E. Garcia, E. Grassia, P. Calyam, and K. A. Hoque, ``PEARL: An adaptive and explainable hardware Trojan detection using open source and enterprise large language models,'' In \emph{IEEE Access}, vol. 13, pp. 133755--133772, 2025.

\bibitem{Yang2025HTDetection}
B. Li, C. Dong, D. Qiu, M. Chen, and Y. Yang, ``Defense in the reverse fragment: RL-based partial netlist hardware Trojan detection,'' In \emph {IEEE/ACM International Conference On Computer Aided Design (ICCAD)}, pp. 1--7, 2025.

\bibitem{Yu2025HTDetection}
Z. Pan, Z. Shu and X. Yu, ``SAGE: Shapley attention graph network for gate-level Trojan detection and localization,'' In \emph{IEEE Computer Society Annual Symposium on VLSI (ISVLSI)}, pp. 1--6, 2025.

\bibitem{Hong2026HTDetection}
W. Hu, B. Li, L. Wu, Y. Li, X. Li, and L. Hong, ``Design for assurance: Employing functional verification tools for thwarting hardware Trojan threat in 3PIPs,'' In \emph{IEEE Transactions on Dependable and Secure Computing}, vol. 23, no. 1, pp. 132--148, 2026.

\bibitem{kidger2020neural}
P.~Kidger, J.~Morrill, J.~Foster, and T.~Lyons, ``Neural controlled differential equations for irregular time series,'' In \emph{International Conference on Neural Information Processing Systems (NIPS)}, pp. 6696--6707, 2020.
  
\bibitem{ma2024presilicon}
P.~Ma, Z.~Wang, and Y.~Wang, ``A pre-silicon detection based on deep learning model for hardware Trojans,'' In \emph{Journal of Circuits Systems and Computers}, vol.~33, no.~8, 2024.

\bibitem{puschner2023redteam}
E. Puschner, T. Moos, S. Becker, C. Kison, A. Moradi, and C. Paar, ``Red team vs. blue team: A real-world hardware Trojan detection case study across four modern CMOS technology generations,'' In \emph{IEEE Symposium on Security and Privacy (SP)}, pp. 56--74, 2023.

\bibitem{mosavirik2023silicon}
T.~Mosavirik, S.~K.~Monfared, M.~S.~Safa, and S.~Tajik, ``Silicon echoes: Non-invasive Trojan and tamper detection using frequency-selective impedance analysis,'' In \emph{Cryptology ePrint Archive}, Paper 2023/075, 2023.

\bibitem{tehranipoor2010survey}
M.~Tehranipoor and F.~Koushanfar, ``A survey of hardware Trojan taxonomy and detection,'' In \emph{IEEE Design \& Test of Computers}, vol.~27, no.~1, pp.~10--25, 2010.

\bibitem{jin2008hardware}
Y.~Jin and Y.~Makris, ``Hardware Trojan detection using path delay fingerprint,'' In \emph{IEEE International Workshop on Hardware-Oriented Security and Trust (HOST)}, pp. 51--57, 2008.

\bibitem{agrawal2007trojan}
D.~Agrawal, S.~Baktir, D.~Karakoyunlu, P.~Rohatgi, and B.~Sunar, ``Trojan detection using IC fingerprinting,'' In \emph{IEEE Symposium on Security and Privacy (SP)}, pp. 296--310, 2007.

\bibitem{hasegawa2017hardware}
K.~Hasegawa, M.~Oya, M.~Yanagisawa, and N.~Togawa, ``Hardware Trojans classification for gate-level netlists using multi-layer neural networks,'' In \emph{IEEE International Symposium on On-Line Testing and Robust System Design (IOLTS)}, pp. 227--232, 2017.

\bibitem{nasr2024siamese}
A.~Nasr, K.~Mohamed, A.~Elshenawy, and M.~Zaki, ``A Siamese deep learning framework for efficient hardware Trojan detection using power side-channel data,'' In \emph{Scientific Reports}, vol. 14, no. 13013, pp.~1--13, 2024.

\bibitem{Faezi2021HTDetection}
S. Faezi, R. Yasaei, A. Barua, and M. A. A. Faruque, ``Brain-inspired golden chip free hardware Trojan detection,'' In \emph{IEEE Transactions on Information Forensics and Security}, vol. 16, pp. 2697--2708, 2021.

\bibitem{Tehranipour2024HTDetection} D. Saha, S. Tarek, K. Yahyaei, S. K. Saha, J. Zhou, M. Tehranipoor, and F. Farahmandi, ``LLM for SoC security: A paradigm shift,'' In \emph{IEEE Access}, vol. 12, pp. 155498--155521, 2024.

\bibitem{chen2018neural}
R.~T.~Q. Chen, Y.~Rubanova, J.~Bettencourt, and D.~K. Duvenaud, ``Neural ordinary differential equations,'' In \emph{Advances in Neural Information Processing Systems}, pp. 6571--6583, 2018.

\bibitem{morrill2021neural}
J.~Morrill, P.~Kidger, C.~Salvi, J.~Foster, and T.~Lyons, ``Neural rough differential equations for long time series,'' In \emph{International Conference on Machine Learning (PMLR)}, pp. 7829--7838, 2021.

\bibitem{debrouwer2019gru}
E.~De~Brouwer, J.~Simm, A.~Arany, and Y.~Moreau, ``GRU-ODE-Bayes: Continuous modeling of sporadically-observed time series,'' In \emph{Advances in Neural Information Processing Systems (NIPS)}, pp. 7379--7390, 2019.

\bibitem{xiao2016hardware}
K.~Xiao, D.~Forte, Y.~Jin, R.~Karri, S.~Bhunia, and M.~Tehranipoor, ``Hardware Trojans: Lessons learned after one decade of research,'' In \emph{ACM Transactions on Design Automation of Electronic Systems (TODAES)}, vol.~22, no.~1,
  pp. 1--23, 2016.

\bibitem{kocher1999differential}
P.~Kocher, J.~Jaffe, and B.~Jun, ``Differential power analysis,'' In \emph{Annual International Cryptology Conference}, pp. 388--397, 1999.

\bibitem{morrill2021online}
J.~Morrill, P.~Kidger, L.~Yang, and T.~Lyons, ``Neural controlled differential equations for online prediction tasks,'' In \emph{arXiv preprint arXiv:2106.11028}, 2021.

\bibitem{dataset_rozhin1}
R. Yasaei, S. Faezi, and M. A. A. Faruque, ``Hardware Trojan power \& EM side-channel dataset,'' In \emph{IEEE Dataport}, https://dx.doi.org/10.21227/9fwb-8978.

\bibitem{Omidi2024EvasiveHardwareTrojan}
B. Omidi, K. N. Khasawneh, and I. Alouani, ``Evasive hardware Trojan through adversarial power trace,'' arXiv preprint arXiv:2401.02342, 2024.

\end{thebibliography}

\end{document}